\documentclass[a4paper]{jpconf}
\usepackage{graphicx}

\newcommand{\noi}{\noindent}
\newcommand{\dint}{\displaystyle\int}
\newcommand{\fr}{\frac}
\newcommand{\hb}{\hbar}

\newcommand{\om}{\omega}
\newcommand{\Om}{\Omega}
\newcommand{\del}{\delta}
\newcommand{\bg}{\begin{equation}}
\newcommand{\en}{\end{equation}}
\newcommand{\dsum}{\displaystyle\sum}
\newcommand{\nn}{\nonumber}

\newcommand{\lan}{\langle}
\newcommand{\ran}{\rangle}
\newcommand{\Gam}{\Gamma}
\newcommand{\ba}{\begin{eqnarray}}
\newcommand{\ea}{\end{eqnarray}}

\newcommand{\bal}{\begin{align}}
\newcommand{\eal}{\end{align}}

\newcommand{\pa}{\partial}

\newcommand{\til}{\tilde}

\begin{document}

\title{Low-frequency squeezing spectrum of a laser driven polar quantum emitter}

\author{A.V. Soldatov}

\address{Department of Mechanics,V.A. Steklov Mathematical Institute of the Russian Academy of Sciences, 8, Gubkina str.,
Moscow, 119991, Russia}

\ead{soldatov@mi-ras.ru }

\begin{abstract}

It was shown by a study of the incoherent part of the
low-frequency resonance fluorescence spectrum  of the polar
quantum emitter driven by semiclassical external laser field and
damped by non-squeezed vacuum reservoir that the emitted
fluorescence field is squeezed to some degree nevertheless.  As
was also found, a higher degree of squeezing could, in principle,
be achieved by damping the emitter by squeezed vacuum reservoir.

\end{abstract}


\section{Introduction}

Squeezed states of electromagnetic (EM) field are of paramount
importance to theoretical and experimental quantum physics
because, besides other useful features, their statistically
observable properties reveal true non-classical nature of light
\cite{Loudon:1987}. By now, these states have also found important
technological applications in high precision measurements,
spectroscopy, high resolution imaging techniques and optical
communications. As a rule, experimental studies of squeezed states
have been carried out with  macroscopic sources of squeezed light
despite that the possibility of squeezed light generation from a
single two-level quantum emitter in free space was theoretically
predicted long ago \cite{Walls:1981}. But only recently this
prediction was experimentally proved to be the case for
high-frequency resonance fluorescence in a semiconductor two-level
quantum dot due its anomalously large transient dipole moment in
comparison to those found in natural atoms an molecules
\cite{Schulte:2015}. It would be of theoretical as well as
practical interest to find also a single-emitter source of
squeezed low-frequency EM field. In the present study it is shown
that a  polar emitter represented by a simple two-level quantum
system with broken inversion symmetry could play this role.
Actually, violation of this symmetry is common in such natural
systems as polar molecules as well as in artificially manufactured
systems, like quantum dots. Due to this violation,  these systems
possess permanent dipole moments. The cause for the inversion
symmetry violation is different for different systems. For
example,
 in quantum dots the violation is induced by the asymmetry of the
  confining potential of the dot. Therefore, this asymmetry can be
  hugely augmented
artificially in comparison to natural polar molecules, where its
origin is due to  the natural parity mixing of the molecular
states \cite{Kovarskii:1999}. However, in all cases this violation
results in non-equal permanent diagonal dipole matrix elements of
the ground and excited states. To our knowledge, the notion that a
simple two-level quantum system driven by high-frequency classical
EM field can emit EM field of much lower frequency if its dipole
operator possesses permanent non-equal diagonal matrix elements,
was revealed in \cite{Kibis:2009} for the first time. This
phenomenon was further studied thoroughly in
\cite{Soldatov:2016,Soldatov:2017,Bogolyubov:2018} for the case of
a two-level system driven by external EM field and damped by a
dissipative thermal reservoir. The case of interaction with a
broadband squeezed vacuum dissipative reservoir was studied
earlier for weak driving EM field in \cite{BSPEPAN:2020,
BSJP:2020}.

\section{Model Hamiltonian}

In this study we consider a two-level atom with ground state
$|g\rangle$, excited state $|e\rangle$, transition  frequency
$\om_0$ and the electric dipole moment $\hat{\bf d}$, driven by
external classical monochromatic field ${\bf E}(t)={\bf
E}\cos(\om_f t)$ with an amplitude ${\bf E}$ and  frequency
$\om_f$, and also coupled to a reservoir  $B$ made of a plurality
of modes of quantized electromagnetic field  being in the squeezed
vacuum state. It is assumed that the frequency Lamb shift due to
interaction with the reservoir is already incorporated into the
atomic transition frequency $\om_0$. Thus, the model Hamiltonian
reads

\vspace{-0.4cm}

\bg H=H_{S}(t)+\hb\dsum_{k}\om_k b^+(\om_k) b(\om_k)+\dsum_k
\left(g(\om_k)S^+b(\om_k)+g^*(\om_k)
b^+(\om_k)S^-\right).\label{model}
 \en

 \vspace{-0.4cm}

\noi Here $S^+=|e\rangle\langle g|$ and $S^-=|g\rangle\langle e|$
are the usual raising and lowering atomic operators and
$S^z=\fr{1}{2}(|e\rangle\langle e| - |g\rangle\langle g|)$ is the
atomic population inversion operator. The operators $ b(\om_k)$
and $b^+(\om_k)$ are the annihilation and creation operators for
the vacuum modes satisfying the commutation relations

\vspace{-0.4cm}

\bg  [ b(\om), b^+(\om')]=\delta(\om-\om'),\,\, [ b(\om),
b(\om')]=0,\,\, [ b^+(\om), b^+(\om')]=0,  \en

\vspace{-0.2cm}

\noi and the  term

\vspace{-0.5cm}

\bg H_{S}(t)=\hb\om_0 S^z+\fr{\hb}{2}\Om_R(S^-e^{i\om_f
t}+S^+e^{-i\om_f t})
 +\fr{\hb}{2}(e^{i\om_f t}+e^{-i\om_f
t})\left[ \delta_a S^z - \fr{\delta_s}{2} (|e\rangle\langle e|+
|g\rangle\langle g|)\right]\label{sf1}
 \en

\vspace{-0.3cm}

\noi contains an interaction between the driving field and the
atom in the rotating wave approximation (RWA). Here $\Om_R=-\bf E
d_{eg}/\hb$ is the Rabi frequency being made real and positive by
the appropriate choice of the phase factors of the states
$|e\rangle$ and $|g\rangle$, and ${\bf d}_{eg}=e\langle e|\hat
{\bf r}|g\rangle,\quad {\bf d}_{ge}=e\langle g|\hat {\bf
r}|e\rangle,\quad {\bf d}_{ee}=e\langle e|\hat {\bf
r}|e\rangle,\quad {\bf d}_{gg}=e\langle g|\hat {\bf r}|g\rangle$
are the atomic dipole moment operator matrix elements. As a rule,
it is assumed that $ {\bf d}_{ee}={\bf d}_{gg}=0$, because typical
physical systems, like atoms and molecules, possess the inversion
symmetry, and each of the states $|g\rangle$ and $|e\rangle$ is
either symmetric or antisymmetric. Contrary to this view, we
assume that the inversion symmetry of the system in question is
violated, $ {\bf d}_{ee}\ne{\bf d}_{gg}$, so that
  $\delta_a={\bf E}({\bf d}_{gg}-{\bf d}_{ee})/\hb$ and $\delta_s={\bf E}({\bf d}_{gg}+{\bf
 d}_{ee})/\hb$. The term proportional to $\delta_s$ does not influence the dynamics of the system
 and can be omitted, while the term proportional to the symmetry violation
 parameter $\delta_a$ is retained.  The squeezed vacuum reservoir source is assumed to be broadband,
and the squeezed vacuum field is characterized by the following
correlation functions \cite{Gardiner:1986,Puri:2001}:

\vspace{-0.5cm}

 \bg  \lan b^+(\om_k)
b(\om_{k'})\ran_{svac}=N(r)\delta(\om_k-\om_{k'}),\,\,\,\lan
b(\om_k)\, b(\om_{k'})\ran_{svac}=
-M(r,\theta)\delta(\om_k+\om_{k'}-2\om_s),\en

\vspace{-0.7 cm}

\bg\lan b(\om_k)
b^+(\om_{k'})\ran_{svac}\!=\!(N(r)+\!1)\delta(\om_k\!-\!\om_{k'}),\lan
b^+(\om_k') b^+(\om_{k})\ran_{svac}\!=\!
-M^*(r,\theta)\delta(\om_{k'}+\om_{k}\!-\!2\om_s),\en

\noi where $\om_s$ is the carrier frequency of the squeezed field,
$r$ is the degree of squeezing, $\theta$ is the phase of
squeezing, $N(r)=\sinh^2(r)$ is related to the mean number of
photons and $M(r,\theta)=\cosh(r)\sinh(r)\exp(i\theta)$ is
characteristic of the squeezed vacuum field and describes the
correlation between the two photons created in the down-conversion
process.

%%%%%%%%%%%%%%%%%%%%%%%%%%%%%%%%%%%%%%%%%%%%%%%%%%%%%%%%%%%%%%%

\section{Equations of Motion for Atomic Variables}

In what follows, it is assumed that $\delta_a \ll \Om_R$, so that
the interaction of the driving field with the permanent dipole
moment is much weaker than its interaction with the transitional
dipole moment. It is also assumed that the driving field itself is
weak.   In the Markoff approximation the master equation for the
atomic reduced density operator $\rho^{rf}_S(t)= e^{i\om_f S^z
t}\rho_S(t) e^{-i\om_f S^z t}$ can be written in the frame
rotating with the driving field frequency $\om_f$ as

\vspace{-0.4cm}

\ba \fr{\pa \rho^{rf}_S(t)}{\pa t}= i\Gamma\delta [S^z,
\rho^{rf}_S(t)]-\fr{i}{2}\delta_a(e^{i\om_f t}+e^{-i\om_f t})
[S^z, \rho^{rf}_S(t)]+ \nn \\
+ \fr{1}{2}\Gam N(r)(2S^+ \rho^{rf}_S(t)S^- -S^-S^+
\rho^{rf}_S(t)-  \rho^{rf}_S(t)S^-S^+)+ \nn\\
+ \fr{1}{2}\Gam ( N(r)+1)(2S^- \rho^{rf}_S(t)S^+ -S^+S^-
\rho^{rf}_S(t)-  \rho^{rf}_S(t)S^+S^-) - \nn\\
 -\Gam M(r,\theta) S^+  \rho^{rf}_S(t)S^+ - \Gam M^*(r,\theta)S^-
\rho^{rf}_S(t)S^-
  -\fr{1}{2}i\Om_R[S^++S^-,  \rho^{rf}_S(t)], \label{master}\ea

 \noi under the
assumption that the carrier frequency $\om_s$ of the squeezed
field coincides with the frequency $\om_f$. Here $\Gamma$ is the
radiative damping constant, $\delta=(\om_f-\om_0)/\Gamma$. A
closed set of equations follows from Eq.(\ref{master}):

\vspace{-0.4cm}

\bg \!\fr{d\lan \til S^-(t)\ran}{dt}\! =\! -\Gam
\left(\fr{1}{2}\!+\!
N(r)\!-\!i\del+i\fr{\delta_a}{2\Gam}(e^{i\om_f t} \!+e^{-i\om_f
t})\right)\!\lan \til S^-(t)\ran+\Gam M(r,\theta)\lan \til S^+(t)
\ran\! +\Om_R\lan S^z(t) \ran, \label{em1}\en

\vspace{-0.9cm}

 \bg \!\fr{d\lan \til S^+(t) \ran }{dt}\!=\! -\Gam
\left(\fr{1}{2}\!+\!
N(r)\!+\!i\del-i\fr{\delta_a}{2\Gam}(e^{i\om_f t} \!+\!e^{-i\om_f
t})\right)\!\lan \til S^+(t)\ran+\Gam\tilde M^*(r,\theta)\lan \til
S^-(t) \ran \!+\Om_R\lan S^z(t)\ran,\label{em2}\en

\vspace{-0.4cm}

\bg\fr{d\lan S^z(t) \ran}{dt} = -\fr{1}{2}\Om_R\left(\lan
S^-(t)\ran+ \lan S^+(t) \ran\right) -\Gam(2 N(r)+1)\lan S^z(t)\ran
-\Gam/2,\label{em3}\en

\noi where $\lan \tilde S^\pm(t)\ran =\pm i \lan S^\pm(t)e^{\mp
i\om_f t }\ran$ are slowly varying parts of the atomic operators.
 The system of equations (\ref{em1}-\ref{em3}) can be solved
numerically  by means of the technique employed earlier in
\cite{Ficek:2001}, where the components of the vector $\vec
X(t)=(\lan \tilde S^-(t)\ran, \lan \tilde S^+(t)\ran, \lan
S^z(t)\ran)$ are decomposed as $
X_i(t)=\dsum_{l=-\infty}^{+\infty}X_i^{(l)}(t)e^{il\om_f t},\,
i=1,2,3, $ and the slowly varying amplitudes $X_i^{(l)}(t)$ obey
the system of equations

\vspace{-0.7cm}

\ba \fr{d}{dt} X_1^{(l)}(t)= -\Gam \left(\fr{1}{2}+ N(r)-i\del
+il\fr{\om_f}{\Gam} \right) X_1^{(l)}(t) -\nonumber\ea

\vspace{-0.7cm}

 \ba
-i\fr{\delta_a}{2}(X_1^{(l-1)}(t)+X_1^{(l+1)}(t))+\Gam M(r,\theta)
X_2^{(l)}(t)+\Om_RX_3^{(l)}(t), \label{emx1}\ea

\vspace{-0.7cm}

\ba \fr{d}{dt} X_2^{(l)}(t)=-\Gam \left(\fr{1}{2}+N(r)-i\del
+il\fr{\om_f}{\Gam} \right) X_2^{(l)}(t) + \nn\ea

\vspace{-0.7cm}

\ba +i\fr{\delta_a}{2}(X_2^{(l-1)}(t)+X_2^{(l+1)}(t))+\Gam
M^*(r,\theta) X_1^{(l)}(t)+\Om_R X_3^{(l)}(t), \label{emx2}\ea

\vspace{-0.7cm}

\ba \fr{d}{dt} X_3^{(l)}(t) =-\fr{\Gam}{2}\delta_{l,0}-(\Gam(2
N(r)+1)+il\om_f)X_3^{(l)}(t)-\fr{\Om_R}{2}
(X_1^{(l)}(t)+X_2^{(l)}(t)). \label{emx3}\ea

\section{Low-frequency squeezing spectrum}

The incoherent part of the fluorescence spectrum can be broken
down into three contributions \cite{Swain:1996, Carmichael:1987}

\vspace{-0.7cm}

\bg F_{inc}(\om) = F_{X}(\om)+F_{Y}(\om)+F_{as}(\om), \en

\vspace{-0.7cm}

\ba F_{X}(\om)=\fr{\Gamma}{2\pi}\mbox{Re}\dint_0^\infty d\tau
\lim_{t\to\infty}\left[\lan \tilde S^+(t)\tilde
S^-(t+\tau)\ran-\lan \tilde S^+(t)\ran\tilde\lan
S^-(t+\tau)\ran\right.+\nn\\
 \left.+\lan\tilde S^+(t)\tilde
S^+(t+\tau)\ran-\lan \tilde S^+(t)\ran\lan\tilde S^+(t+\tau) \ran
\right] \cos\Big((\om-\om_f)\tau\Big),\label{fspecx}\ea

\vspace{-0.7cm}

\ba F_{Y}(\om)=\fr{\Gamma}{2\pi}\mbox{Re}\dint_0^\infty d\tau
\lim_{t\to\infty}\left[\lan \tilde S^+(t)\tilde
S^-(t+\tau)\ran-\lan \tilde S^+(t)\ran\lan\tilde
S^-(t+\tau)\ran\right.-\nn\\
 \left.-\lan\tilde  S^+(t)\tilde
S^+(t+\tau)\ran+\lan \tilde S^+(t)\ran\lan\tilde S^+(t+\tau) \ran
\right] \cos\Big((\om-\om_f)\tau\Big),\label{fspecx}\ea

\vspace{-0.7cm}

\bg F_{as}(\om)=-\fr{\Gamma}{\pi}\mbox{Im}\dint_0^\infty d\tau
\lim_{t\to\infty}\left[\lan \tilde S^+(t)\tilde
S^-(t+\tau)\ran-\lan \tilde S^+(t)\ran\lan\tilde
S^-(t+\tau)\ran\right]
\sin\Big((\om-\om_f)\tau\Big),\label{fspecy}\en

\noi where $ F_{Y}(\om)$ and $F_{Y}(\om)$ are in-phase and
out-of-phase quadrature components of the noise spectrum, and
$F_{as}(\om)$ is the asymmetric contribution. Because of the
so-called quantum regression hypothesis
\cite{Puri:2001,Carmichael:1993}, the fluctuation correlation
functions $ Y_1(t,t+\tau)=\lan \tilde S^+(t)\tilde
S^-(t+\tau)\ran-\lan \tilde S^+(t)\ran\lan\tilde
S^-(t+\tau)\ran,$\,$Y_2(t,t+\tau)=\lan \tilde S^+(t)\tilde
S^+(t+\tau)\ran-\lan \tilde S^+(t)\ran \lan \tilde
S^+(t+\tau)\ran,$\,\, $Y_3(t,t+\tau)=\lan \tilde S^+(t)\tilde
S^z(t+\tau)\ran -\lan \tilde S^+(t)\ran\lan\tilde S^z(t+\tau)\ran,
$ satisfy virtually the same set of equations of motion
(\ref{em1}-\ref{em3}) for the correspondent averages $\lan\tilde
S^-(\tau)\ran$, $\lan\tilde S^+(\tau)\ran$ and $\lan\tilde
S^z(\tau)\ran$ with the only difference that the inhomogeneity
$-\Gamma/2$ disappears due to the subtraction of the mean. These
correlation functions can be decomposed  as
$Y_i(t,t+\tau)=\dsum_{l=-\infty}^{+\infty}Y_i^{(l)}(t,\tau)e^{il\om_f(t+
\tau)}$, $i=1,2,3,$ so that

\vspace{-0.4cm}

\ba \fr{d}{d\tau} Y_1^{(l)}(t,\tau)= -\Gam \left(\fr{1}{2}+
N(r)-i\del +il\fr{\om_f}{\Gam} \right) Y_1^{(l)}(t,\tau)
-\nonumber\ea

\vspace{-0.9cm}

 \ba
-i\fr{\delta_a}{2}(Y_1^{(l-1)}(t,\tau)+Y_1^{(l+1)}(t,\tau))+\Gam
M(r,\theta) Y_2^{(l)}(t,\tau)+\Om_R Y_3^{(l)}(t,\tau),
\label{emy1}\ea

\vspace{-0.8cm}

\ba \fr{d}{d\tau} Y_2^{(l)}(t,\tau)=-\Gam \left(\fr{1}{2}+
N(r)-i\del +il\fr{\om_f}{\Gam} \right) Y_2^{(l)}(t,\tau) + \nn\ea

\vspace{-0.7cm}

\ba
+i\fr{\delta_a}{2}(Y_2^{(l-1)}(t,\tau)+Y_2^{(l+1)}(t,\tau))+\Gam
M^*(r,\theta) Y_1^{(l)}(t,\tau)+\Om_R Y_3^{(l)}(t,\tau),
\label{emy2}\ea

\vspace{-0.9cm}

\ba \fr{d}{d\tau} Y_3^{(l)}(t,\tau) =-(\Gam(2
N(r)+1)+il\om_f)Y_3^{(l)}(t,\tau)-\fr{\Om_R}{2}
(Y_1^{(l)}(t,\tau)+Y_2^{(l)}(t,\tau)), \label{emy3}\ea

 \noi
and the Laplace transforms $ \bar Y_i^{(l)}(t,z)=\dint_0^\infty
e^{-z\tau}Y_i^{(l)}(t,\tau)d\tau $ will satisfy  the following set
of equations:

\vspace{-0.7cm}

\ba z \bar Y_1^{(l)}(t,z)+\Gam \left(\fr{1}{2}+ N(r)-i\del
+il\fr{\om_f}{\Gam} \right) \bar Y_1^{(l)}(t,z) +\nn \\
  +i\fr{\delta_a}{2}\left(\bar Y_1^{(l-1)}(t,z)+\bar
Y_1^{(l+1)}(t,z)\right)  -\Gam  M(r,\theta) Y_2^{(l)}(t,z)
-\Om_R\bar Y_3^{(l)}(t,z)=\nn\\
=\fr{1}{2}\delta_{l,0}+X_3^{(l)}(t)-\dsum_{r=-\infty}^{\infty}X_1^{(l-r)}(t)X_2^{(r)}(t),
\label{emyz1}\ea

\vspace{-0.5cm}

\ba z \bar Y_2^{(l)}(t,z)+\Gam \left(\fr{1}{2}+ N(r)-i\del
+il\fr{\om_f}{\Gam} \right) \bar Y_2^{(l)}(t,z) - \nn\\
 -i\fr{\delta_a}{2}\left(\bar Y_2^{(l-1)}(t,z)+\bar
Y_2^{(l+1)}(t,z)\right)-\Gam  M^*(r,\theta)
Y_1^{(l)}(t,z)-\Om_R\bar Y_3^{(l)}(t,z)= \nn\\=
-\dsum_{r=-\infty}^{\infty}X_2^{(l-r)}(t)X_2^{(r)}(t),\label{emyz2}\ea
\vspace{-0.7cm}

\ba z \bar Y_3^{(l)}(t,z) +(\Gam(2N(r)+1)+il\om_f)\bar
Y_3^{(l)}(t,z)
 +\fr{\Om_R}{2}(\bar Y_1^{(l)}(t,z)+\bar Y_2^{(l)}(t,z))
  =\nn\\=
-\dsum_{r=-\infty}^{\infty}\left(
\fr{1}{2}\delta_{r,0}+X_3^{(r)}(t)  \right)X_2^{(l-r)}(t).
\label{emyz3}\ea

\vspace{-0.5cm}

 \noi   In the steady state limit $(t\to\infty)$
only the zero-order components $\bar Y_{1,2}^{(0)}(t,z)$
contributes to $F_{inc}(\om)$. Therefore,

\vspace{-0.6cm}

\ba F_{X}(\om)=  \fr{\Gamma}{4\pi}\mbox{Re}\lim_{t\to\infty}\left[
[\bar Y_1^{(0)}(t,z)+ \bar Y_2^{(0)}(t,z)]\Big |_{z=-i(\om-\om_f)}
+[\bar Y_1^{(0)}(t,z)+ \bar Y_2^{(0)}(t,z)]\Big
|_{z=i(\om-\om_f)}\right], \label{fspecx1}\ea

\vspace{-0.6cm}

 \ba F_{Y}(\om)=
\fr{\Gamma}{4\pi}\mbox{Re}\lim_{t\to\infty}\left[ [\bar
Y_1^{(0)}(t,z)- \bar Y_2^{(0)}(t,z)]\Big |_{z=-i(\om-\om_f)}
+[\bar Y_1^{(0)}(t,z)- \bar Y_2^{(0)}(t,z)]\Big
|_{z=i(\om-\om_f)}\right], \label{fspecy1}\ea

\vspace{-0.6cm}

\ba F_{as}(\om)= \fr{\Gamma}{2\pi}\mbox{Re}\lim_{t\to\infty}\left[
\bar Y_1^{(0)}(t,z)\Big |_{z=-i(\om-\om_f)} - \bar
Y_1^{(0)}(t,z)\Big |_{z=i(\om-\om_f)}\right]. \label{fspecy1}\ea

%%%%%%%%%%%%%%%%%%%%%%%%%%%%%%%%%%%%%%%%%%%%%%%%%%%%%%%%%%%%%%%%

\vspace{-0.7cm}

\section{Numerical Results}

\begin{figure}[th]
\begin{minipage}{18.5pc}
\includegraphics[width=18.5pc]{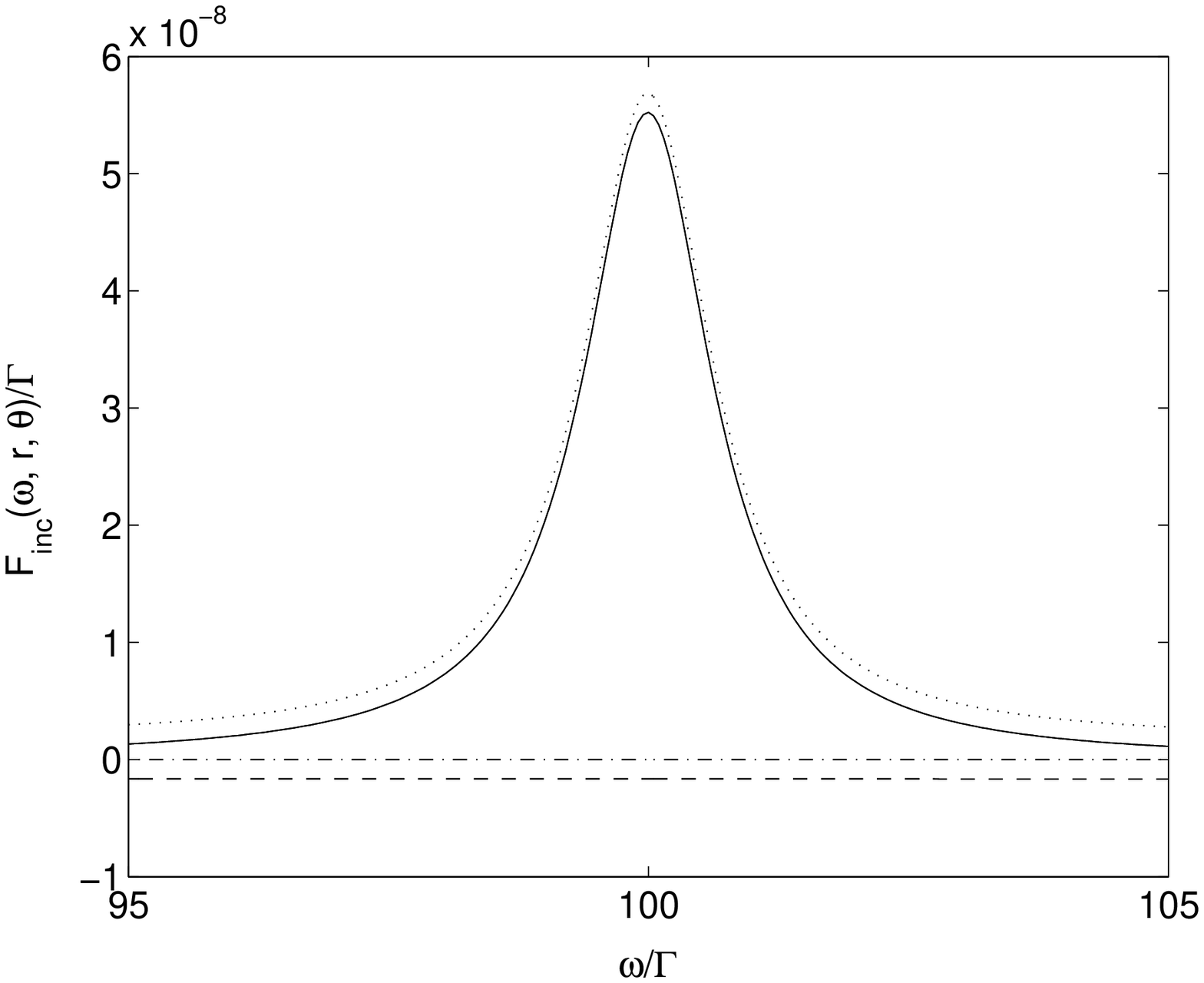}
\caption{\label{fig01} Fluorescence spectrum $F_{inc}$ and its
components  $F_X --$, $F_Y \cdot\cdot$, $F_A -\cdot-$.\\
 $\Gamma=1, \, r=0, \theta=0,\,
\om_f=\om_s=\om_0=5000, \Om_R=100, \delta_a=10$.}
\end{minipage}\hspace{1pc}%
\begin{minipage}{18.5pc}
\includegraphics[width=18.5pc]{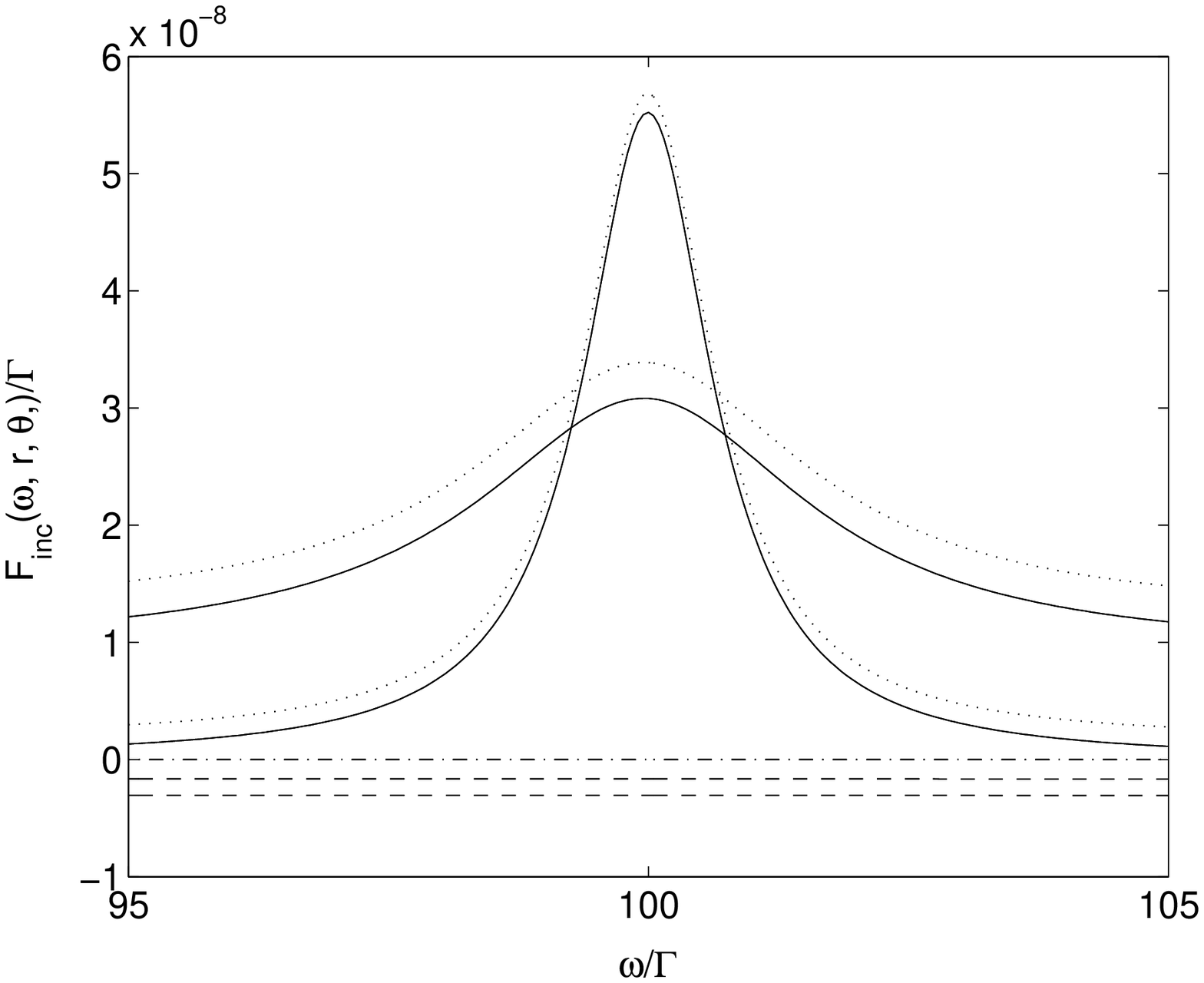}
\caption{\label{fig02}Fluorescence spectrum $F_{inc}$ and its
components  $F_X --$, $F_Y \cdot\cdot$,  $F_A -\cdot-$.\\
$\Gamma=1, \,r=0,\, r=1,\, \theta=0,\, \om_f=\om_s=\om_0=5000,
\Om_R=100, \delta_a=10$.}
\end{minipage}
\end{figure}

\begin{figure}[th]
\begin{minipage}{18.5pc}
\includegraphics[width=18.5pc]{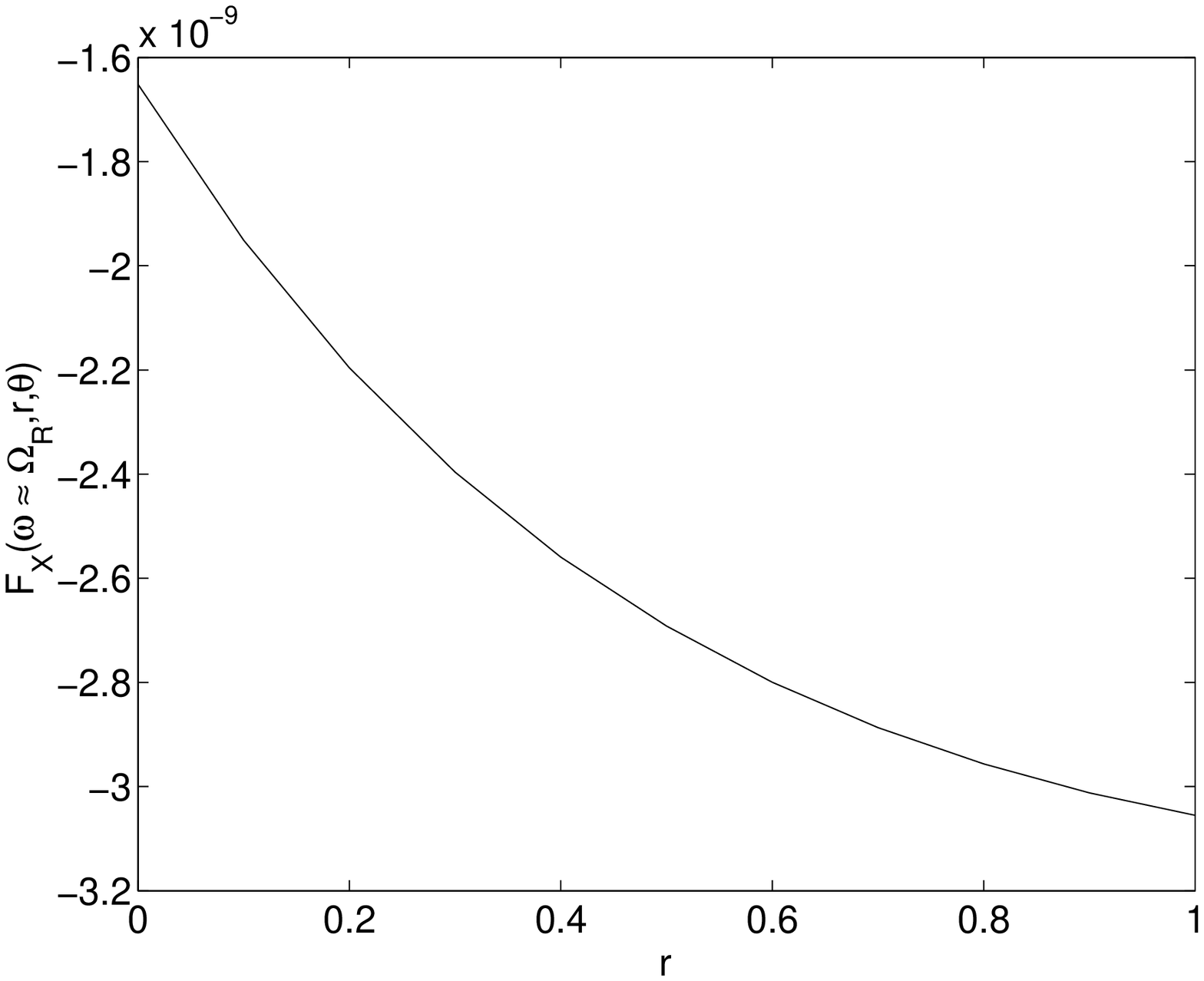}
\caption{\label{fig03}Fluorescence spectrum component $F_X$ at
$\om=\Om_R$ for various values of $r$. \\$\Gamma=1,\, \theta=0,\,
\om_f\!=\!\om_s\!=\!\om_0\!=\!5000, \Om_R\!=\!100,
\delta_a\!=\!10$.}
\end{minipage}\hspace{1pc}%
\begin{minipage}{18.5pc}
\includegraphics[width=18.5pc]{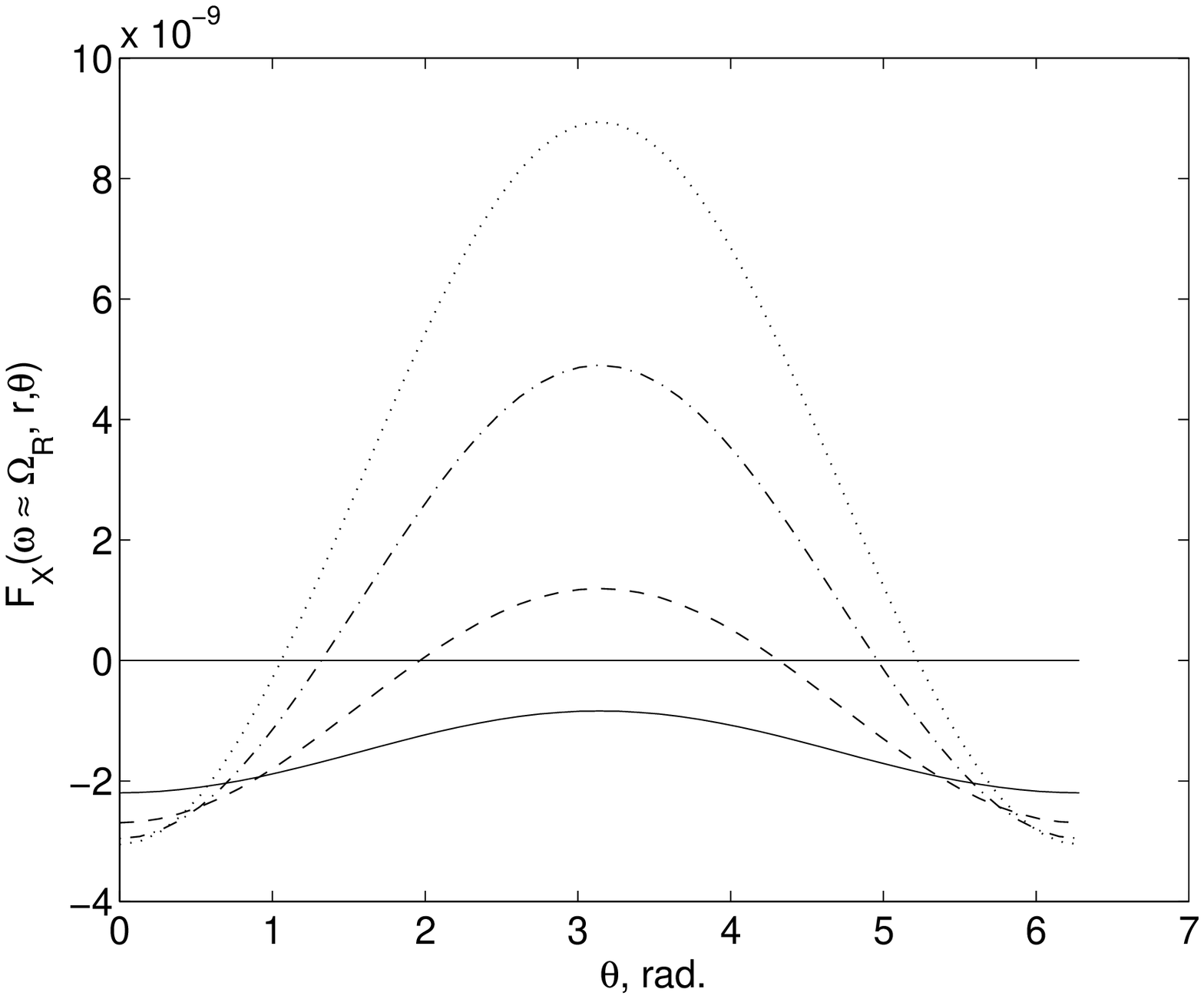}
\caption{\label{fig04}Fluorescence spectrum component $F_X$ at
$\om=\Om_R$ for various values of $r=0.2 -, 0.5 --, 0.8 -\cdot-,
1.0 \cdot\cdot$ and $\theta$. $\Gamma=1, \,
\om_f\!=\!\om_s\!=\!\om_0\!=\!5000, \Om_R\!=\!100,
\delta_a\!=\!10$.}
\end{minipage}
\end{figure}

In this research, the case of the driving field frequency $\om_f$
and the carrier frequency of the squeezed field $\om_s$ being
simultaneously in resonance with the atomic transition frequency
$\om_0$  was studied. Equations (\ref{emx1})-(\ref{emx3}) and
(\ref{emyz1})-(\ref{emyz3}) were solved numerically, as usual
\cite{Ficek:2001}, in the steady state limit $(t\to\infty)$ by
truncation of the number of the harmonic amplitudes $X_i^{(l)}(t)$
and $\bar Y_i^{(l)}(t,z)$ taken into account.  As was shown before
 \cite{Soldatov:2016} for a driven two-level system with
broken symmetry interacting with non-squeezed vacuum reservoir, a
low-frequency radiation peak centered nearly exactly at the
frequency $\om=\Om_R$ appears in the fluorescence spectrum, see
Fig.\ref{fig01}. It is seen that the in-phase quadrature spectral
component $ F_{X}(\om)$ is nearly uniformly negative, which means
that the fluorescent field is squeezed (cf.
\cite{Walls:1981,Collett:1984,Ou:1987,Tanas:1998}) even without
squeezing of the vacuum reservoir field. The increase in the
vacuum squeezing degree $r$ results in the spectral amplitude
decrease and the broadening of the radiation peak at $\om=\Om_R$,
see Fig.\ref{fig02}, while the component $ F_{X}(\om)$ steadily
decreases, see Fig.\ref{fig03}. The degree of the fluorescent
field squeezing is strongly affected by the squeezing phase
$\theta$ of the vacuum field, see Fig.\ref{fig04}. For large
enough $r$, there is a domain of $\theta$ values for which this
squeezing totally disappears.

\vspace{-0.5cm}

\newpage

\vspace{-1.2cm}

\vspace{-0.9cm}

\section{Conclusion}

In conclusion, the effect of the broadband squeezed vacuum
dissipative damping reservoir on  the squeezing properties of the
low-frequency fluorescence field emitted by a quantum two-level
polar system with broken inversion symmetry driven by external
high-frequency classical EM (laser) field  was studied.   As was
found, the squeezing in the low-frequency fluorescent field
already exists even without squeezing in the vacuum field.  It was
also shown that the presence of squeezing in the vacuum field can
increase the degree of squeezing in the fluorescent field for
appropriate values of the vacuum squeezing phase. At the same
time,  it is possible to alternate the amplitude and the spectral
width of the low-frequency fluorescence spectral peak by changing
the parameters of the squeezed vacuum, such as the squeezing
degree $r$ and squeezing phase $\theta$.

\end{document}